\documentstyle[12pt,aps,float,epsfig]{revtex}
\tighten

\begin{document}
\draft

\title{Topological Effects in Ring Polymers:\\
A Computer Simulation Study}

\author{M. M\"{u}ller 
\thanks{Institut f\"ur Physik, Johannes-Gutenberg Universit\"at,
55099 Mainz, Germany},
J. P. Wittmer
\thanks{to whom correspondence should be addressed.}
\thanks{Cavendish Laboratory, Madingley Road, Cambridge CB3 0HE, UK},
M. E. Cates
\\
{\small Dept. of Physics \& Astronomy, University of Edinburgh,} \\
{\small King's Building, Mayfield Road, Edinburgh EH9 3JZ, UK}}
\date{May 1996}
\setcounter{page}{1}
\maketitle

\begin{abstract}
Unconcatenated, unknotted polymer rings in the melt are subject
to strong interactions with neighboring chains due to the presence
of topological constraints. We study this by computer simulation 
using the bond-fluctuation algorithm for chains with up to $N=512$
statistical segments at a volume fraction $\Phi=0.5$ and show
that rings in the melt are more compact than gaussian chains.
A careful finite size analysis
of the average ring size $R \propto N^{\nu}$ yields 
an exponent $\nu \approx 0.39 \pm 0.03$
in agreement with a Flory-like argument for the topological
interactions.
We show (using the same algorithm) that the dynamics of molten rings
is similar to that of linear chains of the same mass, confirming recent
experimental findings. The diffusion constant varies effectively 
as $D_{N} \propto N^{-1.22(3)}$ and is slightly {\em higher} than that of 
corresponding linear chains. 
For the ring sizes considered (up to 256 statistical segments) 
we find only one characteristic 
time scale 
$\tau_{ee} \propto N^{2.0(2)}$; this is shown by the collapse of 
several mean-square displacements and correlation functions
onto corresponding master curves. Because of the shrunken
state of the chain, this scaling is {\it not} compatible
with simple Rouse motion. It applies for all sizes of ring studied and
no sign of a crossover to any entangled regime is found.
\end{abstract}


\section{Introduction}

\begin{figure}[b]
\centerline{\epsfig{file=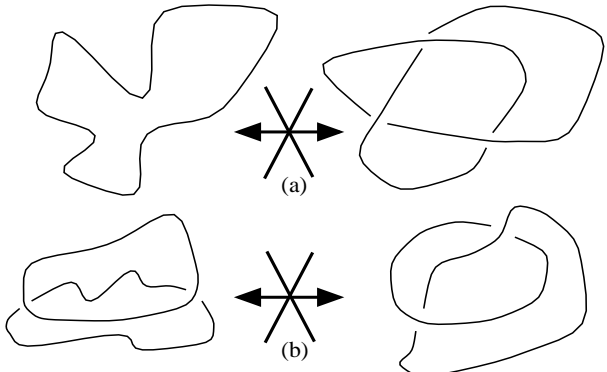,width=80mm,height=50mm,angle=0}}
\caption[]{Sketch of the opposed topological constraints.
a) On the left is a permitted configuration of an unknotted ring.
It cannot turn into the forbidden (knotted) configuration on the right
(nor {\em vice versa}).
b) On the left is a permitted configuration of a pair of unknotted,
unconcatenated rings. It cannot turn into the forbidden (concatenated)
configuration on the right (nor {\em vice versa}).}
\label{figsketch}
\end{figure}

Ring polymers have been extensively studied experimentally
by several groups, not always leading to completely consistent results
\cite{higgins,roovers,hadziioannou,roovers1,mckenna1,mills1,tead1}.
Indeed, the synthesis of the samples of unknotted and nonconcatenated rings 
(see fig.~\ref{figsketch})
remains a very delicate issue \cite{mckenna1}.
Nevertheless, the general conclusion emerging from these studies is 
that the dynamics of rings is quite similar to that of linear chains 
of the same molecular mass and density \cite{mckenna2,tead1}. 
This is certainly a surprising result from the point of view of
the reptation model which describes the flow behavior of
entangled linear polymer chains \cite{doi,lrp}.
In this effectively single-chain model, the constraints on the motion of
a reference chain due to the interactions with its neighbors
are replaced by a curvilinear ``tube" within which the
chain ``reptates": relaxation of the constraints occurs only 
at the chain ends.
Clearly, a closed ring polymer (having no ends) cannot reptate
in this conventional sense. Accordingly, the motion in a melt of 
rings should be quite different from that of an analogous linear chain
system; and at first sight on would expect it to be slower.
Indeed it was argued \cite{klein1,lrp} that motion of ring polymers should be
exponentially slow and comparable with that of star polymers \cite{stars}.

This is however not borne out by rheological measurements
on polystyrene (PS) and polybutadyene (PB) rings, which
showed that the zero-shear viscosities $\eta_{0}$
for melts of rings for all 
molecular weights considered are similar to, but even slightly
smaller than, those for linear chains \cite{roovers,mckenna1}.
The temperature dependence of $\eta_{0}$ in PS rings
is virtually indistinguishable from that of linear polystyrene of
high enough molecular weight; all viscosities display classical WLF or
Vogel type temperature dependence and they can be superimposed on a single
master curve.
For linear chains, two power-law regimes arise in the mass
dependence of the viscosity, corresponding to Rouselike and
entangled motion. Unexpectedly, the same is found for ring polymers.
Below a critical mass $M_{c}$
the viscosity seems to increase with an exponent $\alpha_{1} \approx 3/2$
which is larger than the expected $\alpha_{1}=1$ for ideal Rouse behavior.
The values for rings and linear chains of the same molecular weight are
of similar magnitude.
In the high mass regime ($M>M_c$) the viscosities increase 
more strongly, again with a power similar to the linear case.
An exponential increase in the range of masses available 
(up to $M \approx 1.85 \cdot 10^{5}$) is explicitly 
excluded \cite{mckenna1,viscoexpo}.

The similarity of the dynamics of ring molecules to their linear counterparts is
also reported on a more microscopic level from tracer diffusion measurements
\cite{mills1,tead1}.
Dilute labeled PS chains of mass $M$, in a matrix of mass $P$,
were measured using Forward Recoil Spectroscopy. Different topologies
(rings in linear matrices\cite{mills1}, 
linear chains in ring matrices \cite{tead1},
rings in microgel \cite{tead1}) were investigated and compared to linear chains
in linear matrices. Unfortunately, there exist so far no systematic measurements
of ring tracers in matrices of rings (of identical size) in which the
molecular mass covers a significant range. 
However, consistent with the rheological measurements
quoted above, the tracer diffusion of linear PS in
ring matrices was found to be nearly identical to that of linear PS in
linear PS matrices \cite{tead1,threading}. 
This is again a surprising result in view of
theoretical concepts of matrix dependent tracer diffusion \cite{klein1}.

A suggestion by Lodge et al. \cite{lrp} is that the experiments on rings
may be reconciled
with the reptation concept
by taking into account the higher entanglement length. 
The critical mass for entanglement can be estimated from the viscosities
or from the shear modulus. PS rings exhibit a plateau modulus approximately
one-half (one-fifth for PB) that for linear polymers, 
suggesting that rings are less effective at forming entanglements. 
Rubber elasticity arguments indicate
a critical mass for polystyrene cycles of $M_{c}=58000$ compared to
$M_{c}=30000$ for linear chains. 
If one now makes a comparison {\it at an equal number of entanglements}
ring melts in fact have higher viscosities than linear chains. 
The experimental viscosity data then
lie in the unentangled/entangled crossover region, where it is clear from the
linear polymer results that reptation is not the only available mode.
This explanation would allow a crossover to much slower ring dynamics at
very high molecular weights.

No theoretical explanation for this increase of the critical mass exists,
but it seems reasonable to relate it
to the idea that ring conformations in the melt may be partially
collapsed \cite{mikedeutsch}.
Besides the usual excluded volume interaction between two
neighboring chains there is a topological interaction due
to the exclusion of knotted and concatenated ring configurations 
(see fig.~\ref{figsketch}). 
While the excluded volume interaction is screened out whenever 
the overlap with other chains is large (i.e., as long $R^{d}/N >> 1$,
true for large chains if $\nu > 1/d$) 
the topological interaction in 3-dimensions 
need not be screened and should cause a reduction in the size exponent
$\nu$ below the gaussian value ($\nu = 1/2$).

Despite this argument, no experimental study of the radius of gyration
of rings in the melt appears to have been made.
A good understanding of the static properties of the system is of course
an indispensable starting point for a reasonable description of the
dynamics, so this is unfortunate. 
Detailed static measurements (in both dilute and concentrated systems)
could also provide a stringent test on the quality of the ring
synthesis, as McKenna et al. have pointed out \cite{mckenna2}.
Without them, for many synthesis routes it is hard 
to exclude the possibility of knotted or concatenated rings,
nor that of a small fraction of linear chain contaminant which might
modify the viscosity substantially (by threading the rings, for example)
\cite{mckenna1}.

In view of those experimental difficulties, a detailed simulation
study to investigate both the statics and the dynamics in a melt of
rings is certainly warranted. Most previous attempts are based on
Pakula's ``cooperative motion algorithm'' \cite{pakula1,pakula4}.
While this algorithm can give correct results for the statics 
(especially for melts) its value for studying dynamics is questionable 
(see section 2 below).
In the following we study melts of non-knotted, non-concatenated 
rings within the well-established ``bond-fluctuation model"
(BFM) to get a better insight in their statics and dynamics.
Apart from its computational efficiency, an advantage of the 
method is that comprehensive data for linear chains
have already been obtained \cite{paul,wittmer1} with which quantitative 
comparisons can be made. 
Indeed, even if ring polymers did not exist experimentally, this 
comparison might help illuminate several longstanding but still
controversial issues in the dynamics of entangled linear chains 
\cite{shaffer}.

The article is organized as follows : In the next section we give briefly
some technical comments on the BFM simulation performed. 
In section 3 results on the conformation of isolated rings (in an
athermal solvent) 
are presented, showing that
the unknottedness constraint is insufficient 
to swell the rings significantly beyond what would be caused by excluded
volume forces alone. Molten rings, on the other hand, are found to be 
quite compact with an exponent $\nu \approx 0.4$, consistent with 
the assumption \cite{mikedeutsch} that roughly
one degree of freedom is lost for every topological interaction with a
neighboring chain.
We then show, in section 4, that the dynamics of rings and 
linear chains are qualitatively similar over much of the range of chain length
that we are able to simulate. For the largest rings, however, a
perfect scaling (involving a single characteristic time for each chain
length) is still obeyed whereas for this size of linear chain
significant departures are observed. This allows us to confirm the suggestion that 
the entanglement mass is much larger for rings. 

We also 
study the degree
to which rings thread through one another, a question addressed in 
both sections 3 and 4. This seems to be insufficient to allow
large clusters of mutually entangled material to build up.
The extent of entanglement, quantified roughly as the number
of neighboring rings contacting a given molecule, nevertheless appears
relevant for dynamics: for both rings and their linear counterparts 
the diffusion constant is shown to scale as the
mass of the correlation hole (at least, in the range of chain 
lengths studied).
Finally in section 5 we give our conclusions and
discuss the impact of these ring investigations
on our understanding of entangled polymer dynamics.

\section{A Bond-Fluctuation Model Study}

The algorithm used in this investigation is the well-established
``bond-fluctuation model" (BFM) of K. Kremer et al \cite{bfm}.
This coarse-grained 3-dimensional
lattice model has proven to be especially useful for investigating
the universal features of statics and dynamics in dense polymeric melts
\cite{paul}.
A small number of chemical repeat units (i.e. a Kuhnian segment)
is mapped onto a lattice monomer such that the relevant characteristics
of polymers are retained: connectivity of the monomers along a chain and
excluded volume of the monomers.
Each monomer occupies a whole unit cell
of a simple cubic lattice with periodic boundary conditions.
Adjacent monomers along a polymer are connected via one of 108 allowed 
bond vectors.
These are chosen such that the local 
excluded volume interactions prevents the chains
from crossing each other during their motion. 
This conservation of the topology
ensures that the rings (which are set up initially as thin
``sausages'') remain neither knotted with themselves
nor concatenated with one another during the relaxation and sampling.
In our athermal simulation, ``local jumps'' are realized
by choosing one monomer at random
and attempting to jump over the distance of one lattice
spacing in one of the six basic directions (also randomly chosen).
The attempt is accepted if excluded
volume restriction are satisfied and the new bond vectors to
the neighbors along the ring belong to the allowed set.

There has already been made a simulation study of ring polymers by
Frisch et al \cite{FRISCH} in the framework of the BFM. However, due
to the large CPU-time demands, that study was restricted to single
rings; also the dynamical properties were not very directly addressed.
The only existing simulation data on the statics and dynamics 
of rings in the melt use Pakula's ``Cooperative Rearrangement Algorithm"
\cite{pakula2,pakula3}. In contrast to the local jumps utilized in the
BFM, this algorithm changes
simultaneously and collectively the monomer positions on a 
number of different
chains. While giving correctly the static properties, a clear
correspondence between Monte-Carlo time and real time is yet to be
established for this algorithm and its dynamical interpretation is
accordingly unclear.

In the present BFM investigation we want to extend
the careful study of linear chains made by Paul et al \cite{paul}
to ring polymers and compare our results to their data on linear chains.
At a filling fraction $\Phi=0.5$ of occupied lattice sites, many static
and dynamic features of molten polymeric materials are reproduced by the BFM. 
For example, the single chain conformations 
obey gaussian statistics down to the screening length
$\xi \approx 6$ (in units of the lattice constant) 
of the excluded volume interaction obtained from the
static structure factor \cite{paul}. 
There are extensive 
results on the dynamical properties covering the range from an
unentangled behavior for short chain lengths up to the onset of reptation-like
motion for chain length $N=200$.
Of course, the dynamics
of long polymers in a dense melt slows down dramatically
with growing chain length and therefore poses huge demands
on CPU-time requirements. 
For the present investigation we employ a very efficient implementation
of the BFM on a massively parallel CRAY T3D \cite{T3D}. 
Using a 2 dimensional geometrical decomposition
of the simulation grid 
of linear extension $L=128$, we employ 64 T3D processors. 
This permits us to equilibrate
systems comprising 131072 monomers and 
ring lengths up to 512 (statics) or 256 (dynamics) statistical segments.
This study involved about $5000$ hours of single processor CPU time.

The starting configurations consisted of straight
sausage-like rings (loops enclosing no area) 
which were carefully equilibrated for 
at least one relaxation time (i.e. the center of mass had 
moved a distance comparable to the chain size) before any
data were taken. Indeed, for all but the longest chains, runs were
continued well beyond this (so as to generate dynamical data)
which enabled us to confirm that the static chain extensions 
had settled to their equilibrium values by this time.

\section{Statics : Conformations of Rings in the Melt}

In this section we consider the effects of the
unknottedness and nonconcatenation constraints on the
conformational properties of dilute and molten rings. 
While the effect of the former constraint on isolated 
rings turns out to be irrelevant the non-concatenation requirement 
significantly compactifies molten rings.  

The size of the rings is measured firstly with the usual
mean-square radius of gyration $\left<R_{g}^{2}\right>$; as a second measure,
we define the average distance
between pairs of monomers that are $N/2$ monomers apart along 
the ring contour,
$\left< R_{e}^{2} \right> = 
\left< \left(\vec{R}_{n}-\vec{R}_{n+N/2}\right)^{2} \right> $,
and call this the mean-square ring diameter. 
For isolated rings in an athermal solvent results for both quantities
are presented in table~\ref{tab1}. 
In agreement with previous Monte Carlo studies by Frisch et al \cite{FRISCH}
the simulation yields a ring size $R \propto N^{\nu}$ with exponent
$\nu \approx 0.595$ for the radius of gyration
and $\nu \approx 0.605$ for the ring diameter. These values are 
only slightly larger
than the excluded volume exponent ($\nu = 0.588$) for linear 
athermal chains, the difference lying within the 
range of the statistical error. 
The influence of the topological 
constraints on the static properties of isolated rings thus appears not to
alter the chain-swelling exponent in 3 dimensions 
(though there may be a prefactor effect) \cite{nud2}. 
This finding concurs with the analytical studies of des Cloizeaux \cite{SR}
and is now also confirmed experimentally. 
Indeed, while early SANS data of J.~Higgins et al \cite{higgins}
seemed to indicate a nearly gaussian statistics for rings in good solvent,
later the delicate dependency of the ring properties on preparation
conditions was overcome by G.~Hadziioannou et al \cite{hadziioannou}
and J.~Roovers et al \cite{roovers}
corroborating that the statistics are as for linear chains in a good solvent. 
However, a decrease of the $\theta$-temperature of isolated rings (compared to
linear chains) by several Kelvin has been measured. This is defined by
the point at which the second virial coefficient from the
light scattering measurements equals zero, and seems to be the only
manifestation of the unknottedness constraint
\cite{roovers,mckenna1}. A Monte Carlo study of the $\theta$-point
depression of isolated rings, including a careful finite-size analysis could
certainly yield interesting additional information; we do not attempt
this here.

\begin{figure}
\centerline{\epsfig{file=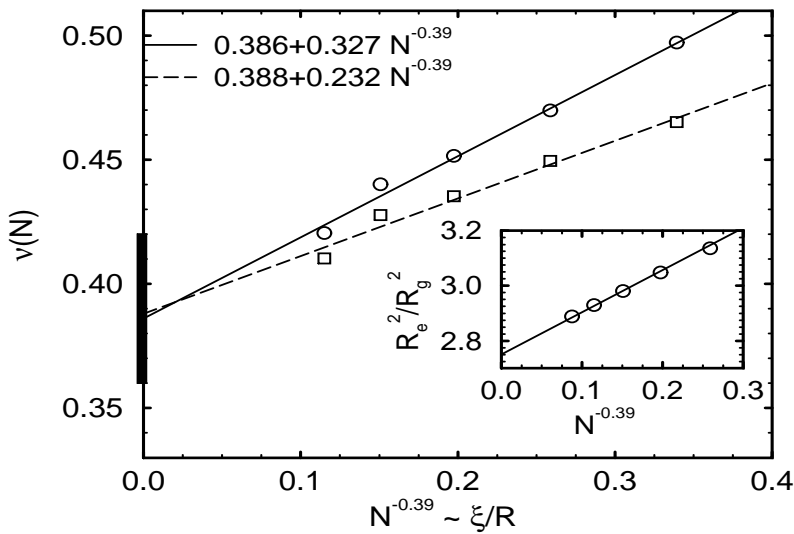,width=85mm,height=120mm,angle=-90}}
\caption[]{The running exponents $\nu(N)$ obtained from the radius
of gyration $\left< R_{g}^{2} \right>$ (circles) and the ring diameter
$\left< R_{e}^{2} \right>$ (squares) versus the natural variable
of the screening $N^{-\nu}$.
Both the two exponents and the ratio 
$\left< R_{e}^{2} \right>/\left< R_{g}^{2} \right>$ 
in the inset lay on straight lines.
From this finite size analysis
we obtain the exponent $\nu \approx 0.39 \pm 0.03$ in the limit of
infinite mass.}
\label{figfsa}
\end{figure}

The situation changes completely in the other limit of molten rings 
where, confirming Pakula's simulation \cite{pakula1} we find
very compact rings (table~\ref{tab2}).
When naively fitted without any finite size analysis, the ring sizes
yield an exponent $\nu \approx 0.44$ for the ring diameter $R_{e}$
and a slightly higher value, $\nu \approx 0.45$, 
for the radius of gyration $R_{g}$.
Those values are quite similar to the ones obtained by Pakula.
However, finite-size corrections are clearly detactable 
in the curvature of the data points, and a more careful fitting is
necessary,
as explained below and in 
fig.~\ref{figfsa}. This yields in an extrapolation to the limit of 
infinite masses a distinctly lower exponent $\nu=0.39 \pm 0.03$, which
(as it should) becomes the same for our two measures of ring size, $R_e$
and $R_g$.

At the volume fraction $\Phi = 0.5$ used in the simulation, 
the excluded volume interaction is not
completely screened, as pointed out in the previous section.
This explains why the radius of gyration, which is more sensitive
to short-scale structure, tends to give larger
exponents $\nu(N)$ than the diameter which probes larger 
distances (fig.~\ref{figfsa}).
At high masses this short-scale effect of the excluded volume
will become less and less important, so both the ratio of diameter to the
radius of gyration (inset of fig.~\ref{figfsa}), and the running
exponents $\nu(N)$,
tend to become smaller. 
These exponents are plotted
against the natural variable of the screening 
(proportional to $\xi/R_g$) \cite{CASASSA}; 
the procedure brings all the measured
values $\nu(N)$ to lie on straight lines which permit a precise finite
size scaling analysis \cite{funcanaly}.
As one sees from fig.~\ref{figfsa} the low masses gives exponents
close to those obtained previously; 
the highest masses used yield an exponent $\nu \approx 0.42$.
Only in the limit of infinite masses does the short-scale 
excluded volume effect vanish and the two lines for $R_{e}$ and $R_{g}$ 
merge at a value $\nu = 0.39 \pm 0.03$.

This value is consistent with a crude but interesting Flory-like estimate
of the free energy of a polymer ring given by Cates and Deutsch 
\cite{mikedeutsch}. They argued that,
if a ring of polymerization index $N$ has size $R$, it is overlapped with
a number of neighboring rings of order $R^{d}/N$ (in $d$ dimensions).
The more spatially extended the ring, the more entropy is lost by
the nonconcatenation constraint with its neighbors. The simplest possible
estimate of this is to say that 
the number of degrees of freedom lost due to the constraint is
proportional to  the number $R^{d}/N$ of neighbors 
which the ring is prevented from
threading. This gives a contribution to the free energy of
$F \propto kT R^{d}/N$ tending to decrease the ring size. On the other
hand, there is also an entropy penalty if the ring becomes too squashed;
the free energy required to squash a gaussian chain of $N$ steps
into a region of linear size $R$ less than $N^{1/2}$ scales as $kT N/R^{2}$.
Adding these contributions, and minimizing over $R$, gives
a characteristic size scaling as $R \propto N^{\nu}$ with $\nu = 2/(d+2)$. 
Hence in three dimensions the exponent is $\nu = 2/5$, very close
to the value found by our simulation. 
Note that the latter value is 
definitely smaller than the result $\nu = 1/2 - 1/6\pi$ put forward
recently by M.~Brereton and T.~Vilgis 
\cite{breretonvilgis1}.

\begin{figure}[b]
\centerline{\epsfig{file=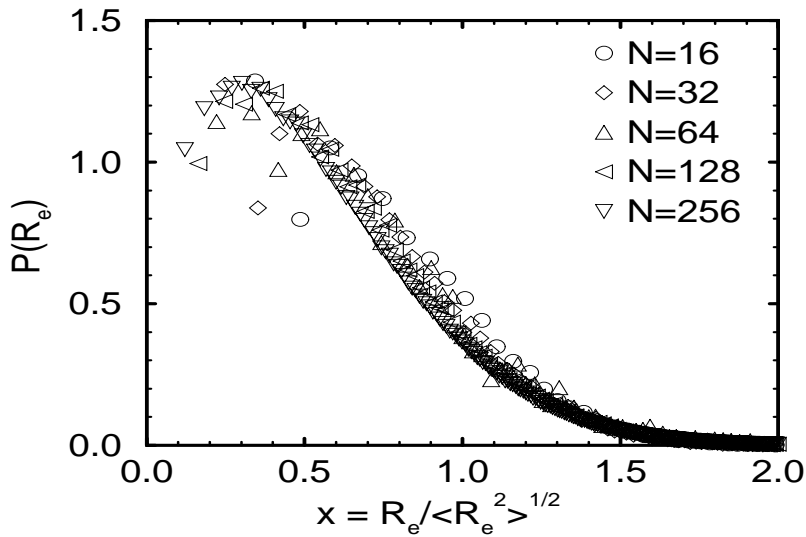,width=120mm,height=80mm,angle=0}}
\caption[]{Distribution function of the ring diameter $R_{e}$ as
function of the reduced variable $R_{e}/\left< R_{e}^{2} \right>^{1/2}$.
Data for various masses are distinguished by different symbols,
as indicated in the figure.
Apart from the very small chains all distributions superimpose.}
\label{figprobRe}
\end{figure}

We note that from the measured exponents of the chain size of
an isolated chain $\nu_{0}\approx 0.6$ 
and of ring chains $\nu_{m}\approx 0.4$ a crossover scaling 
can be obtained in the usual way. This yields for the ring size in the
semidilute
concentration range,
$R \approx R_{0} \left(\Phi/\Phi^{*}\right)^{- 1/4}$
where the monomer density is $\Phi$
and the usual crossover density (overlap threshold) is $\Phi^{*}$.
The decrease in the size of a ring with $\Phi$ is much more pronounced
than for linear chains where $R\sim \Phi^{-1/8}$. 

In fig.~\ref{figprobRe}, we show simulation data for the 
probability distribution $P(R_{e})$ of the diameter of a ring as
a function of the characteristic variable 
$x=R_{e}/\langle R_{e}^2\rangle^{1/2}$.
Data is given 
for various masses up to $N=256$.
The scaling collapse becomes more and more perfect with increasing mass
indicating again that the local interactions become irrelevant
for ring sizes much larger than 
the excluded volume screening length $\xi$.
In principle, this distribution
should define some further characteristic exponents, assuming that, as
for linear chains \cite{degennes} the distribution rises with a
power $P(x) \propto x^{g}$  for small $x$ and drops off
essentially as  $P(x) \propto x^{a} \exp(-x^{\delta})$ for large $x$.
For isolated linear chains the exponents $g$ and $\delta$ can be written
in terms of $\nu$ and $\gamma$ (the latter is the exponent controlling
the $N$ dependence of the free energy).
Unfortunately, although the scaling in fig.~\ref{figprobRe}
is good, the data is not precise enough to extract any corresponding 
exponents in this case. Accordingly we leave this issue for
future investigations.

\begin{figure}
\centerline{\epsfig{file=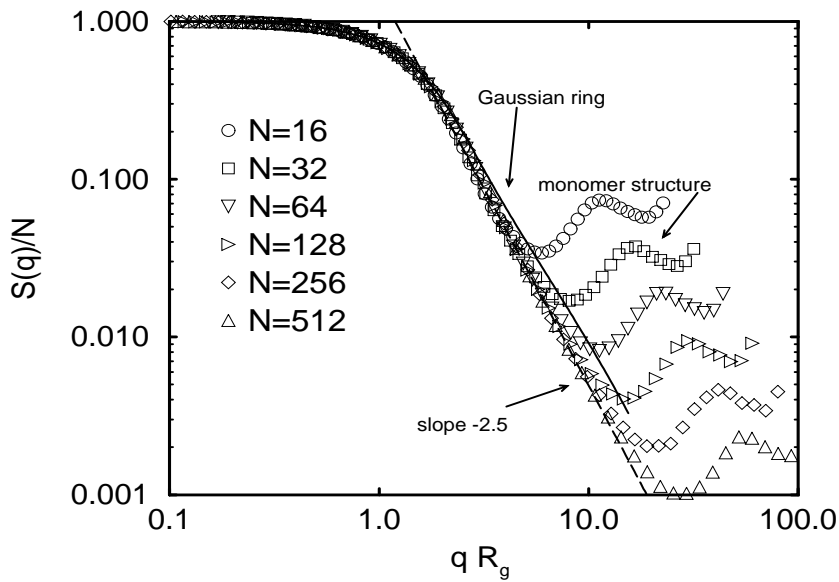,width=90mm,height=120mm,angle=-90}}
\caption[]{Structure factor $S(q)/N$ versus the characteristic variable
$R_{g} q$ for various masses $N$ as indicated in the figure. 
The dashed line confirms the fractal dimension $1/\nu = 2.5 = 1/0.4$.
The solid slope 2 for rings with gaussian statistics cannot match the 
measured structure factor.} 
\label{figSq}
\end{figure}

A quantity of more direct experimental relevance than $P(r)$  is
the static structure factor $S(q)$. 
In fig.~\ref{figSq} the values of $S(q)/N$ for masses up to $N=512$,
expressed as function of the characteristic variable $R_{g} q$,
superimpose (apart from the spurious Bragg peak in each case)
onto a single master curve.
Consistently with our earlier discussion of the scaling of the ring size, 
a self-similar power-law regime is apparent, and shows a fractal dimension
$1/\nu = 1/0.4$ as expected. The Gaussian dimension 2 is clearly ruled
out by our data \cite{gaussring}.

\begin{figure}
\centerline{\epsfig{file=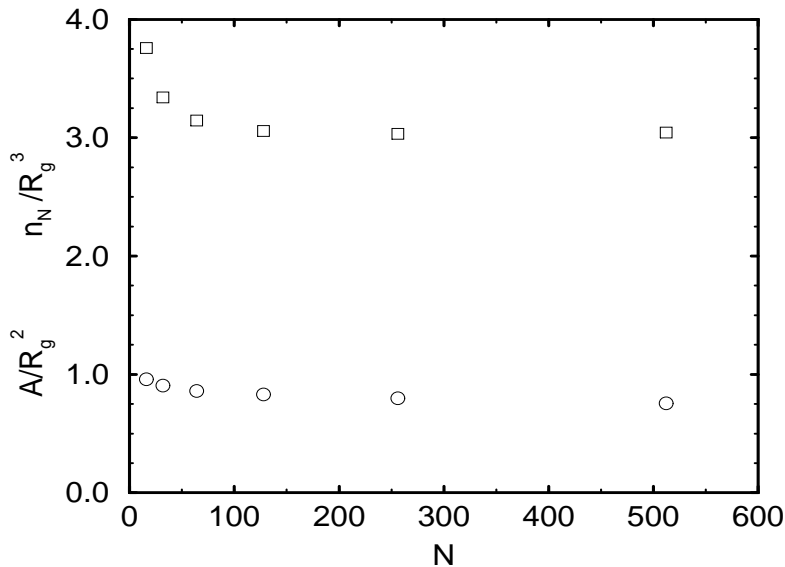,width=90mm,height=120mm,angle=-90}}
\caption[]{Surface circumscribed by a ring in the melt $A$, plotted as
$A/R_{g}^{2}$ 
(circles) and the number of neighbors $n_{N}$, 
plotted as $n_{N} N/R_{g}^{3}$ (squares) versus $N$.
The mass-independent plateaus 0.8 and 3 respectively are reached 
for masses larger than $N=100$.}
\label{figA}
\end{figure}

As mentioned in the introduction, and above, the probability of
threading of a ring polymer by its neighbors is implicated in the
mechanism of partial collapse, and may also have dynamical consequences
by way of long-lived clusters of entangled material 
slowing down the relaxation times. 
It is difficult to define precisely what is meant by threading
(in either static or dynamical terms) but a partial measure of the ease
with which a ring can be threaded is to measure the area of its
projection onto a random direction. This area $a$ is defined as a
signed quantity (the component in that direction of the vector 
area of the ring) which vanishes for any configuration in which the
ring exactly retraces its own steps. A measure of the ``threadability"
is provided by $A = \langle|a|\rangle$; 
it turns out that $A \propto R_{g}^{2}$
as is shown in fig.~\ref{figA}.
The scaling is the same as would be naively expected.
Note that sausage-like configurations would involve very small values 
of $A$. 
The magnitude $A/R_g^2 \approx 1$
indicates that the rings have no tendency to retrace their own steps; 
a fact also revealed by inspection of snapshots.

Another quantity that may be relevant in measuring the 
strength of
the topological interactions is the number of chains touching
a reference chain $n_{N}$. 
Two chains are defined to be ``touching" 
whenever their monomers include pairs separated by a  
distance less or equal to $\sqrt{6}$ lattice constants.
(This somewhat arbitrary microscopic distance
was chosen to include all the monomers within 
the first peak of the monomer-monomer correlation function.)
Intuitively $n_N$ should vary as the number of chains
within the ``correlation hole" spanned by a given chain, and we
see in fig.~\ref{figA} that $n_{N}$ indeed  scales as $R_{g}^{3}/N$.
The plateau at high enough masses confirms the
self-similarity of rings in the melt which was already indicated 
by the scale invariance of the diameter distribution 
in fig.~\ref{figprobRe}.

\section{Dynamics of Rings in the Melt}

As pointed out in the introduction, there is some experimental evidence
that the dynamics of rings are similar to their linear counterparts,
at least up to the largest molecular masses that can readily be obtained.
Our simulation data, presented below, confirms this for rings of 
up to $N=256$ monomers by comparing mean-square displacements, 
chain and cooperative motion correlation functions,
and the resulting diffusion coefficients and relaxation times.

\begin{figure}
\centerline{\epsfig{file=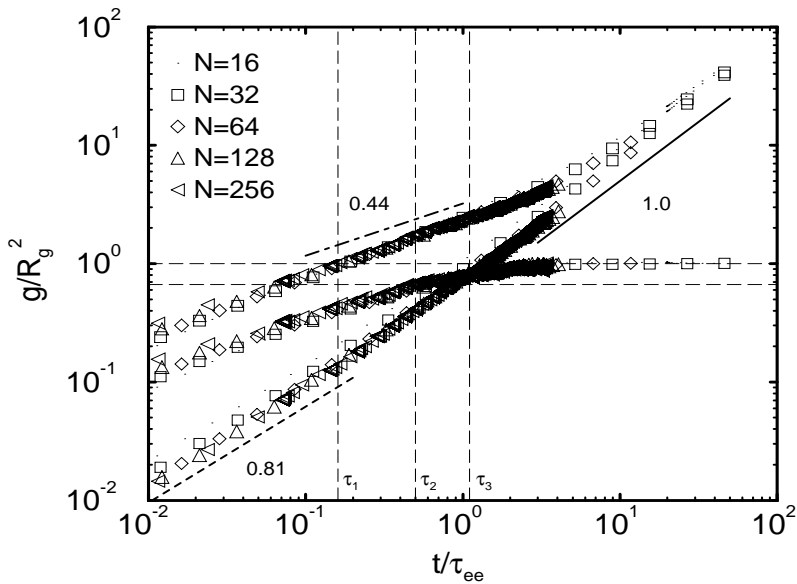,width=90mm,height=120mm,angle=-90}}
\caption[]{Mean-square displacements of a ring monomer $g_{1}(t)$,
of a monomer in the frame of the center of mass $g_{2}(t)$ 
and of the center of mass $g_{3}(t)$ versus the reduced time
$t/\tau_{ee}$.
The three times $\tau_{1,2,3}$ (vertical dashed lines) are defined in the text.
The two horizontal dashed lines correspond to mean-square displacements
of $2/3 \left< R_{g}^{2} \right>$ and $\left< R_{g}^{2} \right>$ respectively.
The Fickian behavior at long times is indicated by the solid line.
The effective exponent $x=0.81$ of the center of mass motion
at shorter times due to net forces of neighboring chains
is displayed by the broken line.
The anomalous diffusion of a monomer for molten rings
with exponent $\nu = 0.4$ is indicated by the dashed-dotted line.
}
\label{figg1g2g3}
\end{figure}

\begin{figure}
\centerline{\epsfig{file=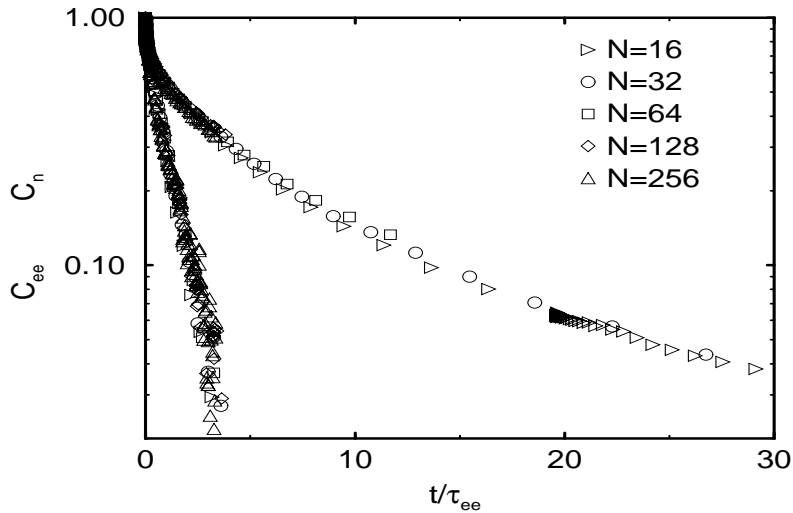,width=80mm,height=120mm,angle=-90}}
\caption[]{
While the correlation function of the diameter vector $C_{ee}(t)$
(data on the left side of the figure) drops off exponentially,
the correlation function $C_{n}(t)$ measuring the mean
number of chains touching a given reference chain decays much
more softly with a power law.
Data points for various masses superimpose when plotted versus
the reduced time $t/\tau_{ee}$. 
}
\label{figCn}
\end{figure}

\begin{figure}
\centerline{\epsfig{file=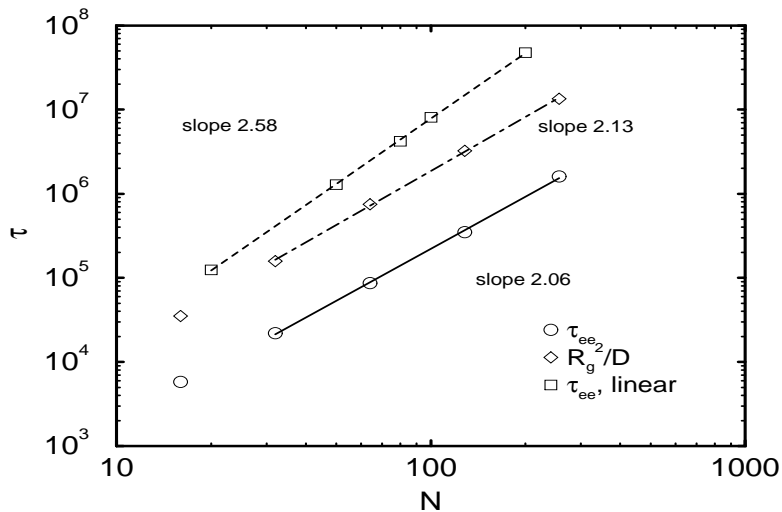,width=80mm,height=120mm,angle=-90}}
\caption[]{
Reorientation relaxation time $\tau_{ee}$ for rings (circles)
and linear polymers (squares) obtained from $C_{ee}(t)$
versus polymer mass $N$. 
The solid line indicate the exponent 2.06 found for the collapsed
molten rings. 
This value is also confirmed (dashed-dotted line) by the 
the relaxation time $R_{g}^{2}/D_{N}$ for rings (diamonds).
The relaxation times of linear chains 
can be fitted by a a power law with exponent 2.58 as indicated
by the dashed line.
}
\label{figtauee}
\end{figure}

We characterize the dynamics by measuring three different 
mean-square displacement functions, describing the
motion of monomers $g_{1}(t) = 
\langle({\bf R}_n(t)-{\bf R}_n(0))^2\rangle$ in the lab frame, 
the motion of monomers in
the center-of-mass
frame of a given ring, 
$g_{2}(t) = \langle ({\bf R}_n(t) -
{\bf R}_{cm}(t) - {\bf R}_n(0) + {\bf R}_{cm}(0))^2\rangle/2$, 
and the motion of the center of mass itself, $g_{3}(t)=
\langle ({\bf R}_{cm}(t)-{\bf R}_{cm}(0))^2\rangle$. 
The mean-square displacement functions for molten rings are
given in fig.~\ref{figg1g2g3}. These quantities are scaled by the 
mean square
characteristic ring size $\langle R_{g}^{2}\rangle$ 
whereas the time
coordinate is scaled by the rotational relaxation time $\tau_{ee}$.
This rotational time is the decay time for relaxation of a
ring diameter
(the vectorial displacement between monomers $N/2$ apart in the sequence)
which is obtained from the correlation function $C_{ee}(t)$
described below (shown in fig.~\ref{figCn}), and its scaling with $N$
is shown in fig.~(\ref{figtauee}).
Apart from the mass $N=16$ all curves perfectly superimpose
with the use of this single scale factor for each chain. 
This indicates that no second time-scale is present
as would be expected for an entangled system (the entanglement time). 
Such scaling is sometimes called Rouse behaviour \cite{paul},
but here we describe it
as ``unentangled scaling" since the power laws involved need not
be those of the Rouse model.
Notably, no crossover whatever is detectable from this scaling to
a modified (reptation or other) motion, involving a second time scale, 
even for the
largest rings studied ($N=256$).
This contrasts with the fact that for linear chains of mass larger than
$N=200$ the unentangled scaling 
starts to break down, with clear signs of a crossover to a new
regime (whose exponents could not, however, be reliably measured) 
\cite{wittmer1}.
This finding confirms that any entanglement length for rings is
larger than that for linear chains of the same type,
in agreement with experimental inference \cite{roovers1,mckenna2}.

As mentioned before, it is
tempting to associate this with the fact that the rings are
partially collapsed. Crudely one could argue that for a given
degree of entanglement, the number of chains in the correlation
hole (which scales as $R_{g}^{3}/N$) should be the same for rings
and linear chains. This would give a rough estimate of
$N = 1000$ BFM monomers as the breakdown point of the unentangled scaling in
the ring case. This estimate is broadly consistent with the entanglement
mass for rings reported in experiments as lying 
between 2 (for PS) or 5 (for PB)  times as large as for linear chains, 
as discussed in the introduction.

Qualitatively the mean square displacements of rings displayed
here show very similar behavior to that of short linear chains, 
up to masses of order $N=100$ \cite{paul,wittmer1}. 
(However as mentioned above, for larger
masses the scaling breaks down in the linear chain case,
and in this regime the rings and linear chains are no longer 
precisely alike.)
While the center of mass follows
the Fickian type of diffusion at long times ($t/\tau_{ee} \gg 1$)
a clear signature of net forces acting on the center of mass of the chains 
is displayed at short times, when the center of mass displacement $g_{3}$
is proportional to $(t/\tau_{ee})^{x}$, with an effective
exponent $x\approx 0.81$. (Within a true Rouse model, in which each
monomer in the system
is subject to uncorrelated random forces, this exponent
would have the Fickian value of unity at all times.)
This effect was also observed for linear chains with
a slightly higher effective exponent $x\approx 0.85$ at the same density 
$\Phi=0.5$ \cite{paul,wittmer1}. 
There it was also verified that the exponent $x$ is density dependent,
approaching 1 in the low density limit;
it is very likely that this is similar for rings.
We can therefore conclude that, as for linear chains, 
there is a net force acting on the center of mass generated
by the interaction of the test chain with surrounding ones,
slowing down the chain motion at short times.

The monomer motion may also be split into two regimes characteristic
of short and long times.
While in the latter the monomeric displacements approach 
asymptotically that of the whole chain ($g_{1}(t)\propto t$),
the conformational properties of the rings in the melt are
reflected in the short time anomalous diffusion regime.
The involvement of more and more ring monomers as the lifetime
of a fluctuation increases, gives by 
general scaling arguments a mean-square monomer displacement of 
$g_{1} \propto (t/\tau_{ee})^{1/(1+1/2\nu)}$.
For our rings ($\nu \approx 0.4$) this yields an exponent of about 0.45
which agrees well with the dynamical
simulation data shown in fig.~\ref{figg1g2g3}. 
(The agreement is
even better if one uses for each chain length the measured 
exponent $\nu(N)$ which is somewhat larger than the extrapolated
value $\nu(\infty)$,
as explained in the previous section.) 
Defining a monomeric mobility $W$ by
$g_{1} = b^2(Wt)^{1/(1+1/2\nu)}$ where $b$ is the monomer
size, perhaps surprisingly we find (as shown in tab.~\ref{tab2} ) 
that $W \approx 7 \cdot 10^{-3}$ for rings is about 4 times
larger than for linear chains where $W \approx 1.6 \cdot 10^{-3}$. 
The mean-square monomer displacement in the frame of the 
center of mass, $g_{2}$,
follows (as expected) for small times the same behavior
of that in the lab frame
$g_{1}$ while for long times it approaches (by definition) 
the mean-square radius of gyration of the rings.

In order to characterize succinctly
the mean square displacement curves for rings, and to compare them
to the linear chain counterparts, we define (following ref.~\cite{paul})
three characteristic times, $\tau_{1,2,3}$ according to the
following criteria: $g_{1}(\tau_{1})=\left< R_{g}^{2} \right>$,
$g_{2}(\tau_{2})= 2/3  \left< R_{g}^{2} \right>$ and
$g_{2}(\tau_{3})=g_{3}(\tau_{3})$. Each $\tau$-parameter is a 
measure of the decay of the corresponding displacement function, as
indicated in fig.~\ref{figg1g2g3}.
Apart from a bit of scatter we obtain, independently of mass,
the ratios $\tau_{2}/\tau_{1} \approx 3.3$ 
(compared to $\approx 0.9$ for linear chains) and  
$\tau_{3}/\tau_{1} \approx 8$ (compared to $\approx 3$).
This means that, normalizing by the motion of a monomer in the lab
frame,
both the center of mass motion, and that of monomers in the
center of mass frame, take
longer to reach their long time asymptotic limits than is the case
for linear chains.

\begin{figure}
\centerline{\epsfig{file=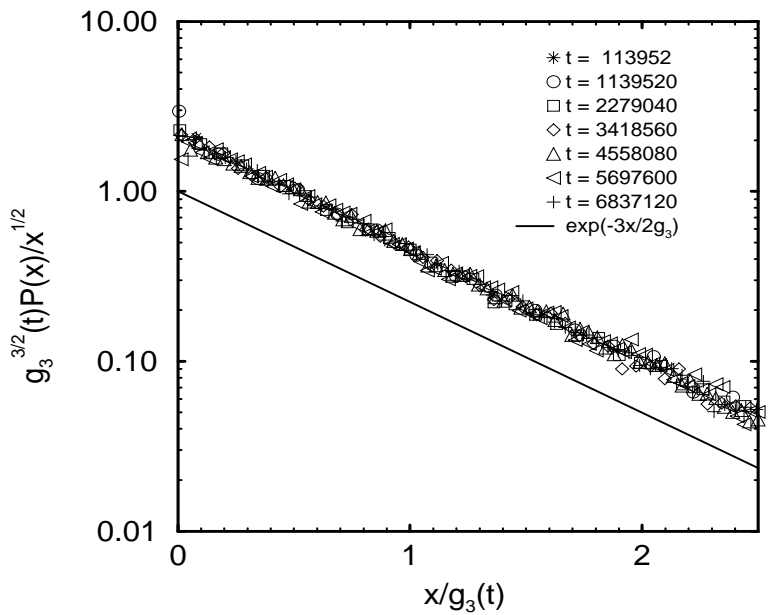,width=90mm,height=120mm,angle=-90}}
\vspace*{0.5cm}
\caption[]{Distribution $P(x)$ of the mean-square displacements $x$ 
of the center of mass $g_{3}(t)$ for chains of mass $N=256$.
Data over a large range of time (as indicated in the figure) 
collapse perfectly on one straight line with slope $-3/2$
when plotted as $g_3^{3/2}(t) P(x)/x^{1/2}$ versus $x/g_{3}(t)$,
corresponding to a gaussian distribution of the displacement variable.
}
\label{figdistg3}
\end{figure}

We now discuss the {\it probability distribution} 
$P(x)$ for the mean square center of mass
displacement $x = ({\bf R}_{cm}(t) - {\bf R}_{cm}(0))^2$.  
In linear chain systems, all chains behave roughly alike and,
on the time scale of motion over one or more gyration radii, the
distribution of ${\bf R}_{cm}$ is essentially gaussian. 
This would lead to 
$P(x) \propto g_3(t)^{-3/2} x^{-1/2}\exp[-3x/2g_3(t)]$.
For rings, another scenario is
possible, in which at a given time a small number of rings are
relatively ``unentangled" (for example with wormlike configurations),
as discussed by Klein \cite{klein1}, while others form entangled
clusters which can scarcely move. This would yield, in the crudest
picture, a bimodal distribution for $P(x)$ at times shorter than the
lifetime of a cluster.
This possibility is apparently ruled out by the distributions $P(x)$
we obtained as for instance in fig.~\ref{figdistg3} 
for a melt of rings of mass $N=256$.
The distribution for times ranging from much smaller than 
the relaxation time $\tau_{ee}=1.6 \cdot 10^6 MCS$,
up to times much larger, could be superimposed on one single master
curve, which shows precisely the 
form expected for gaussian scaling as considered above.

\begin{figure}
\centerline{\epsfig{file=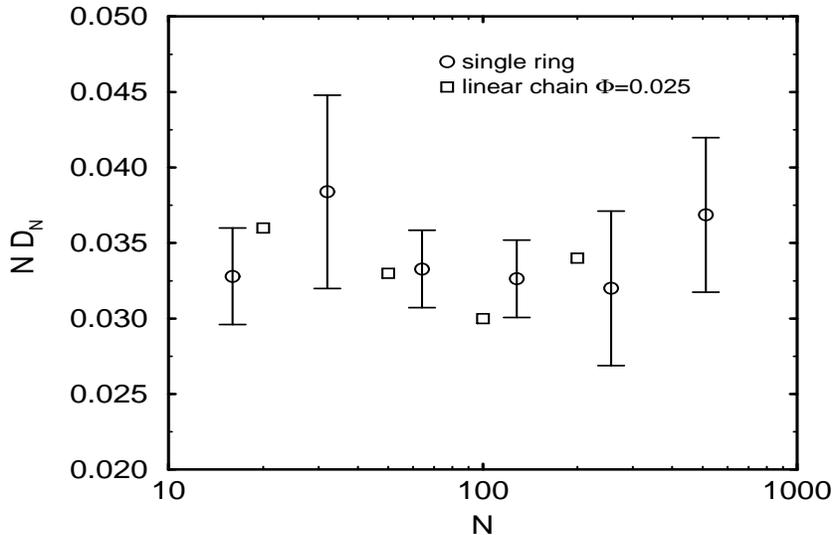,width=80mm,height=120mm,angle=-90}}
\vspace*{0.5cm}
\caption[]{Diffusion coefficient $N D_{N}$ versus mass $N$ 
for isolated rings (circles) and linear chains (squares).}
\label{figdiff_single}
\end{figure}

\begin{figure}
\centerline{\epsfig{file=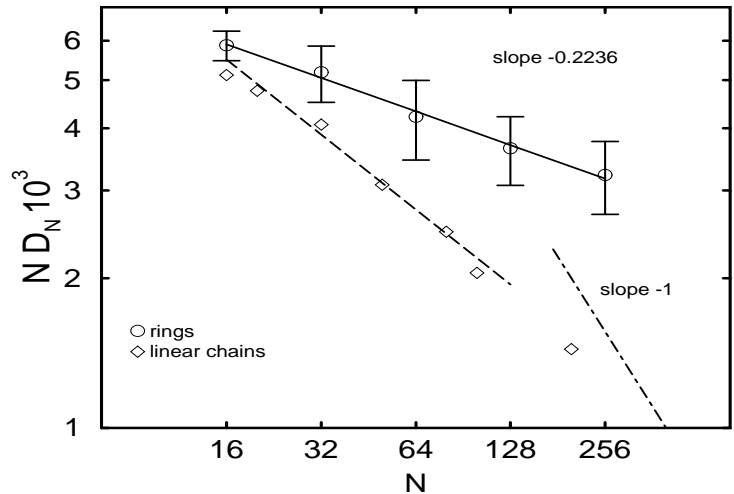,width=80mm,height=120mm,angle=-90}}
\caption[]{Diffusion constant of rings (circles) 
and linear polymer (squares) as a function of the mass $N$.
The diffusion constant of rings varies effectively as 
$N D_{N} \propto N^{-0.22}$ (solid line).
Neglecting any curvature of the data points the diffusion constants
of linear chains are shown (dashed line) to vary as $N D_{N} \propto N^{0.5}$.
This is compared to the reptation prediction for linear chains 
(dashed-dotted line). 
} 
\label{figdiff}
\end{figure}

From the mean-square displacement of the center of mass $g_{3}(t)$
one obtains the diffusion coefficients $D_{N}$ as shown
in fig.~\ref{figdiff_single} for isolated rings and in fig.~\ref{figdiff} 
for rings at a volume fraction $\Phi=0.5$.
For {\it isolated} rings in athermal conditions, 
the diffusion constants obey $D_{N}\approx 0.034/N$; these 
are virtually identical to their linear counterparts 
(fig.~\ref{figdiff_single}), as one expects from an essentially Rouse
dynamics in this limit (hydrodynamic forces are of course excluded
in our model).
The situation is somewhat different for the 
diffusion constants $D_{N}$ obtained for rings in the melt. 
The first striking point is that the rings diffuse {\it faster} than
linear chains of same mass and density. This confirms the trend 
found experimentally in zero-shear viscosities
for PS and PB melts of rings \cite{mckenna2}.

Second, for rings, there is no sign of curvature on the log
plot, which is consistent with the idea that
the entanglement mass for rings is larger than for linear
chains. In fact, the diffusion constant of linear chains, also shown in 
fig.~\ref{figdiff}, has previously been interpreted in terms of a crossover
from true Rouse behaviour for small $N$ (which would appear as
a plateau on this
representation) to a new, entangled
regime, with the crossover effects first appearing for $100\le N \le 200$. 
For linear chains (but not rings) this crossover behaviour was much more 
strongly seen in $g_1$ and $g_3$. Looking at the 
center of mass diffusion data alone, however,
there is no sign of a small-$N$ plateau, and no 
firm evidence of a crossover to a new regime of entangled behaviour,
even in the linear chain case! This suggests that other interpretations 
might also be worth investigating.

If one assumes that, within the range of masses studied, both linear chains
and rings follow {\it a single} power law behavior, then 
one finds respectively $D\sim N^{-1.5}$ and $D\sim N^{-1.2}$ for the
two cases \cite{sameexpon}. 
Remarkably therefore, in each
case that the diffusion constant varies inversely 
with the mass of the ``correlation hole'' of surrounding
chains: $D_{N} \propto 1/R_g^{3}$.
This suggests a mental picture in which the center-of-mass
mobility of any given chain is governed by having to ``drag" the contents
of the correlation hole along with it. 
Although this picture
should clearly break down for long linear chains (where
a more conventional reptation picture becomes appropriate)
it could provide some insight into the behaviour at intermediate $N$.
For rings, this ``intermediate" regime appears to be more strongly
developed. If one compares rings and linear chains at equal 
mean-square
gyration radius, then the
rings have slightly smaller diffusion constants than linear chains. 
Assuming that any increase in the entanglement length of rings
is due to partial collapse, this is consistent with the experimental
fact that the viscosities of ring melts, 
compared with linear chains at an equal number of entanglements, have 
slightly higher viscosities. 

Master curves for some further correlation functions 
are presented in fig.~\ref{figCn}.
The correlation function $C_{ee}(t)$ is 
analogous to the end-end vector correlation
function usually used for linear chains \cite{doi}. 
It describes the decay of the diameter
vector $\langle{\bf R}_e(t)\cdot {\bf R}_e(0)\rangle$
between two monomers of a ring separated by $N/2$ monomers.
The relaxation time $\tau_{ee}$, used to scale the time axis
in the various plots discussed already,
is defined as the time at which this correlation function has 
decayed by a factor $1/e$. Note that this correlation function
shows a near-exponential decay.

This contrasts with the
second correlation function $C_{n}(t)$, which measures 
the decay in the mean
number of chains, ``touching" (in the sense defined above)
a given reference chain
at time zero, that are still touching it (or touching it
again) at time $t$. From this quantity is subtracted an
asymptotic plateau value (arising from the finite size of
the simulation cell) so that the
correlation function vanishes at long time;
we normalize to $C_n(0) = 1$.
This $C_{n}(t)$ apparently decays with a power law (fig.~\ref{figCn}). 
This correlation function $C_{n}(t)$ was defined in an attempt
to monitor any possible 
clustering of entangled rings (leading to a fraction of
slow-moving material), by detecting possible long-lived 
contacts between rings that might arise from threading of one
ring by another.
Although the number of contacting rings remains large for times
much longer than the measured diffusive relaxation times, there is
nothing like a plateau; the power-law decay of
$C_{n}$ apparently scales with $\tau_{ee}$, so there is no new time
scale due to clustering \cite{Cnlinear}.
It seems that the threading probability 
is not large enough to form large clusters whose percolation
(for example) could lead to new dynamical phenomena.

In fig.~\ref{figtauee} we compare the relaxation times $\tau_{ee}$ 
as defined
from the decay of the diameter correlation function $C_{ee}(t)$
(and used above for the scaling of the time axis) with the quantity
$\langle R_g^2\rangle/D$ which also defines a characteristic time.
To a good accuracy, these are proportional, but the latter is larger by
a factor of about 10. The scaling of the relaxation time, by either
definition, follows approximately an $N^2$ law
(as do the times $\tau_{1}$, $\tau_{2}$ and $\tau_{3}$ we obtained from
the mean square displacements), 
as predicted from the simple Rouse model
for gaussian chains or rings. However, we believe that this is 
fortuitous, the diffusion constant being the more fundamental quantity; 
indeed, true Rouse motion for partially collapsed
objects (as we know the rings to be) would give a higher power.
For comparison, the relaxation time obtained from the diffusion
coefficient via $R_{g}^{2}/D_{N}$
for linear chains is also plotted. In the range of $N$ studied,
this follows a $N^{2.6}$ power law.
(The linear chain values are taken out of Ref. \cite{paul},
some data for short chains are added.)

\section{Conclusion}

In this article we have demonstrated using the bond-fluctuation model
various pronounced effects of topological
constraints on the static and dynamic properties of rings.
While the size of isolated rings scale as their linear counterparts, 
molten rings are quite compact objects characterized by an
exponent $\nu=0.39 \pm 0.03$ for the chain size $R$. This is consistent 
with a crude Flory-like argument \cite{mikedeutsch}
suggesting that one degree of 
freedom is lost for
every non-concatenation constraint. 
 
Experimentally, 
melt of rings seem to show dynamical behavior very similar to linear chains,
at least for the masses of rings usually studied \cite{mckenna2}. The same
seems to be true in our computational study, which is of course limited
to relatively modest masses -- though ones for which, in linear chains,
the onset of entanglements would clearly be detectable. 
We see no sign
of a similar effect for rings, suggesting a larger effective
entanglement length in the rings case.
Indeed, in our simulation (which ignores hydrodynamics but respects the
topological
constraints) conformational relaxation of the rings
is well described by a dynamical scaling involving a single
characteristic time scale $\tau_{ee}$ for each chain, whereas an entanglement
crossover would introduce a second timescale. Nonetheless, the motion
is not that of a simple Rouse model (chains moving independently subject
to local friction and uncorrelated noise):
interchain forces are manifest in the reduced effective exponent
for the center of mass motion at short time scales, or
equivalently in an increase in the exponent for the dependence of the
single-chain relaxation time $\tau_{ee}$ on chain length. This
increase, fortuitously, approximately restores the exponent to
the value 2 predicted by a simple Rouse model for gaussian rings,
despite the fact that our rings are partially collapsed.

The breakdown of a true Rouselike dynamics (which would give $g_3(t)\sim
t$) for molten chains is not completely unexpected, since the Rouse model
describes only one chain
or ring in isolation.
What is remarkable from the simulation is that, for 
the range of masses (of both rings and linear chains) studied so far
by BFM methods, the slowing of the center of mass motion is 
consistent with a picture in which each chain or ring has to drag with it 
the neighbors in its correlation hole against the resistance of 
local frictional forces ($D(N) \sim R_g(N)^{-3}$). This very simple
picture may have something to offer for understanding the diffusion
of molten linear polymers in what is classically viewed as the crossover
region between entangled and unentangled motion, although further work
is required to check whether the same relation holds over a range
of volume fractions. 
Note that the diffusion constants and relaxation times
for linear chains obtained with the BFM
agree perfectly with simulations obtained with MD by Kremer
and Grest \cite{paul} and experimental values of Richter by RNA \cite{richter};
these were always analyzed previously in terms of a crossover,
although such data do not show true Rouse behavior even for small $N$ (and
nor does the experimental data unless corrections are made for the
dependence of the effective segmental mobility on chain length 
\cite{ferry}).

Likewise for rings one has to choose whether to assume a crossover or
fit an intermediate power law to the data. The crossover interpretation 
is rather forced for rings, since for the ring sizes
studied here, the intermediate (apparently power-law) behavior shows
no sign of breaking down at either the small or large mass end of the
range. Since the nature of entanglements in ring systems remains
to be clarified, it is possible that this behavior could extend to
quite high masses (or even, in principle, 
be the true limiting result for high $N$). 
Certainly if an entanglement crossover is present
for rings we can say that it is at substantially higher masses than for
linear chains of the same type. 
In the range of $N$ studied, there is in particular no evidence for
a crossover to a regime in which the rings have exponentially long
relaxation times. Such behavior would anyway be surprising since
even rings in a fixed network show an algebraic dependence of the
diffusion constant on mass \cite{mikedeutsch,reiter91}.
In molten linear chains, it is known that reptation cannot be the only 
mode of relaxation (the prefactors predicted from the 
reptation picture for the diffusion coefficient and for the viscosity
systematically underestimate the relaxation); so there are probably enough
alternative modes of motion to allow relaxation of molten rings on an
algebraic timescale even in the limit of high masses. A full
investigation
of that limit must of course await further increases of computer power,
especially since, as emphasized above, any crossover to the entangled
regime occurs for substantially 
higher masses than in the case of linear chains.

\vspace{0.2cm}

{\bf Acknowledgement:}
M.M. thanks K.~Binder, A.~Bruce, W.~Paul and S.~Pawley for 
stimulating discussions, W.Oed for help in porting the 
program to the CRAY T3D, EPFL and EPCC for generous 
access to their CRAY T3D and financial support
by the TRACS-program and DFG grant Bi 314/3-3.
J.W. thanks M.~S.~Turner and W.~Paul for stimulating discussions.


\newpage

\begin{table}
\begin{tabular}{|c|c|c|c|c|} \hline
$N$ & $\left< b^2 \right>$ & $\left< R_g^2 \right>$ & $\left< R_e^2 \right>$ & $D_{N} 10^4$ \\
\hline
\hline
16  & 7.436 & 17.5  & 59.9  &  20.5(20)    \\
32  & 7.455 & 41.1  & 140.2  &  12(2)      \\
64  & 7.464 & 95.9 & 322.9 &   5.2(4)    \\
128 & 7.468 & 221.9 & 739.6 &   2.55(20)    \\
256 & 7.470 & 510   & 1698   &  1.25(20)    \\
512 & 7.472 & 1159  & 3740  &   0.72(10)   \\ \hline
\end{tabular}
\caption{
Mean-square bond length $\left< b^2 \right>$, 
radius of gyration $\left< R_g^2 \right>$, ring diameter $\left< R_e^2 \right>$
and diffusion constant $D_{N}$ for single unknotted rings.
}
\label{tab1}
\end{table}

\begin{table}
\begin{tabular}{|c|c|c|c|c|c|c|c|c|} \hline
$N$ &  $\left< b^2 \right>$ & $\left< R_g^2 \right>$ & $\left< R_e^2 \right>$ & $A$ & $n_{N}$ &  $W 10^{3}$ &  $D_{N} 10^4$    &   $\tau_{ee}/10^4$ \\
\hline
\hline
16  & 6.904 &  12.9 &  42.3 &  12.36 & 10.88 & 5.2 & 3.67(25)   & 0.58        \\
32  & 6.913 &  25.7 &  80.6 &  23.30 & 13.60 & 6.6 & 1.62(21)  & 2.2        \\
64  & 6.920  & 49.3 & 150.3 &  42.46 & 17.01 & 7.3 & 0.66(12)   & 8.7        \\
128 & 6.924 &  92.2 & 274.8 &  76.66 & 21.15 & 7.5 & 0.285(45) &  35        \\
256 & 6.926 & 169.7 & 497.2 & 135.36 & 26.18 & 6.6 & 0.126(21) &  160        \\
512 & 6.927 & 304.0 & 878.0 & 229.88 & 31.50 & & & \\ \hline
\end{tabular}
\caption{
Mean-square bond length $\left< b^2 \right>$,
radius of gyration $\left< R_g^2 \right>$, 
ring diameter $\left< R_e^2 \right>$,
ring surface $A$, number of neighbors touching a reference chain $n_{N}$,
monomer mobility $W$,
diffusion constant $D_{N}$ and rotational relaxation time $\tau_{ee}$
for unknotted, unconcatenated rings in the melt 
at volume fraction $\Phi=0.5$.
}
\label{tab2}
\end{table}
   

\begin{thebibliography}{99}

\bibitem{higgins}
K.~Dogson and J.~Higgins, in {\em Cyclic Polymers}, e.d. by J.~A.~Semlyen,
Elsevier Appl. Sci. Publ. London, N.~Y., 1986.
 
\bibitem{roovers}
J.~Roovers and P.~M.~Toprowski, Macromolecules {\bf 16}, 843 (1983);
J.~Roovers, J.~Pol.~Sci.~Polym.~Phys.~Ed. {\bf 23}, 1117 (1985).

\bibitem{hadziioannou}
G.~ Hadziioannou, Bull. Am. Phys. Soc. {\bf 31}, 355 (1986).

\bibitem{roovers1}
J.~Roovers, Macromolecules {\bf 18}, 1359 (1985).

\bibitem{mckenna1}
G. B. McKenna, G. Hadziioannou, P. Lutz, G. Hild, C. Strazielle,
C. Straupe, P. Rempp and A. J. Kovacs, Macromolecules {\bf 20}, 498 (1987).

\bibitem{mckenna2}
G. B. McKenna, B. J. Hostetter, N. Hadjichristidis, L. J. Fetters and
D. J. Plazek, Macromolecules {\bf 22}, 1834 (1989).

\bibitem{mills1}
P. J. Mills, J. W. Mayer, E. J. Kramer, G. Hadziioannou, P. Lutz,
C. Strazielle,
P. Rempp and A. J. Kovacs, Macromolecules {bf 20}, 513 (1987).

\bibitem{tead1}
S. F. Tead, E. J. Kramer, G. Hadziioannou, M. Antonietti, H. Sillescu,
P. Lutz and C. Strazielle, Macromolecules {\bf 25}, 3942 (1992).

\bibitem{doi}
M. Doi and S. F. Edwards, {\em The Theory of Polymer Dynamics} 
(Clarendon Press, Oxford, 1986).
 
\bibitem{lrp}
T. P. Lodge, N. A. Rotstein and S. Prager, {\em Dynamics of entangled 
polymer liquids: Do linear chains reptate ?}, Advances in Chemical Physics,
Volume LXXIX, 1990.

\bibitem{klein1}
J. Klein, Macromolecules {\bf 19}, 105 (1986).

\bibitem{stars}
J.~Klein, D.~Fletcher and L.~J.~Fetters,  Nature {\bf 304}, 526 (1983);
C.~R.~Bartels, B.~Crist, Jr., L.~J.~Fetters and W.~W.~Graessley,
Macromolecules {\bf 19}, 785 (1986);
M.~Antonietti and H.~Sillescu,  Macromolecules {\bf 19}, 798 (1986).

\bibitem{ferry}
J.~D.~Ferry, {\em Viscoelastic Properties of Polymers},
3rd ed. (Wiley, New York, 1980).

\bibitem{viscoexpo}
If only the data points for the high masses are considered
the exponent in the entangled regime is $\alpha_{2} \approx 3.5$.
If one insists on a Rouse-like behavior ($\alpha_{1}=1$) at low molecular
weight as for linear chains, the viscosities are fitted with 
$\eta_{0} = a M + b M^{\alpha_{2}}$ giving a somewhat higher exponent in
the entangled regime of $\alpha_{2} \approx 3.8 - 3.9$ \cite{mckenna1}.

\bibitem{threading}
The motion of rings in linear matrices is, on the other hand, 
slowed down compared
to linear chains in linear matrices. This is easy to rationalize 
in terms of the threading of rings by linear matrix chains \cite{mills1}.

\bibitem{mikedeutsch}
M.E. Cates and J. M. Deutsch, 
J.~de Physique {\bf 47}, 2121 (1986).

\bibitem{pakula1}
T. Pakula and S. Geyler, 
Macromolecules {\bf 21}, 1665 (1988).

\bibitem{pakula4}
S. Geyler and T. Pakula,
Macromol. Chem., Rapid. Commun. {\bf 9}, 617 (1988).

\bibitem{paul}
W. Paul, K. Binder, D. Heermann and K. Kremer,
J. Phys. II {\bf 1}, 37 (1991);
J.~Chem.~Phys. {\bf 95}, 7726 (1991).

\bibitem{wittmer1}
J. Wittmer, W. Paul and K. Binder, 
Macromolecules {\bf 25}, 7211 (1992).


\bibitem{shaffer}
J.~S.~Shaffer, J.Chem.Phys {\bf 103}, 761 (1995).

\bibitem{bfm}
H.-P. Deutsch and K. Binder, J. Chem. Phys. {\bf 94}, 2294 (1991);
I. Carmesin and K. Kremer, Macromolecules {\bf 21}, 2819 (1988).

\bibitem{FRISCH}
H.L. Frisch, M. Schulz and P.R.S. Sharma, Comp.Poly.Sci. {\bf 4}, 13 (1994).

\bibitem{pakula2}
T. Pakula, Macromolecules {\bf 20}, 679 (1987).

\bibitem{pakula3}
T. Pakula and S. Geyler, Macromolecules {\bf 20}, 2909 (1987).

\bibitem{T3D}
M.~M{\"u}ller, K.~Binder and W.~Oed, Sci. Programming in press;
M.~M{\"u}ller, preprint presented at the {\em Supercomputer '95}, Mannheim.
 
\bibitem{nud2}
The exponent $\nu$ for isolated rings in two dimensions is also similar 
to the linear chain result as shown by several authors 
\cite{gersappe,MFisher} obtaining $\nu \approx 0.745$; in this
case, however, it is not clear that a distinction can be made between
the excluded volume and topological interactions.

\bibitem{gersappe}
D.~Gersappe and M.~Olvera de la Cruz,
Phys.~Rev.~Lett. {\bf 70}, 461 (1993).

\bibitem{MFisher}
M.~E. Fisher, Physica (Amsterdam) {\bf 38 A}, 112 (1989).

\bibitem{SR}
J.~des Cloizeaux and M.~L. Metha, J.~Physique {\bf 40}, 665 (1979).

\bibitem{CASASSA}
E.~F.~Casassa, J. Polym.Sci {\bf A3}, 605 (1965).

\bibitem{funcanaly}
Fortunately the universal function corresponding to
the finite size effect appears to be analytic.

\bibitem{breretonvilgis1}
M.~G. Brereton and T.~A. Vilgis, 
J.~Phys. A, {\bf 28}, 1149 (1995). 

\bibitem{degennes} 
P.-G. de Gennes, {\em Scaling Concepts in Polymer Physics},
Cornell University Press, Ithaca, N. Y., 1979.

\bibitem{gaussring}
For comparison, the structure factor 
for a Gaussian ring is given e.g.\ in ref. \cite{burchard}.
A detailed comparison of molten rings with the Gaussian statistics was
made previously in ref.~\cite{pakula1}.

\bibitem{burchard}
W. Burchard and M. Schmidt, Polymer {\bf 21}, 745 (1980).

\bibitem{sameexpon}
Skolnick et al \cite{Skolnick} also obtained in their
simulation of linear chains at volume fraction $\Phi = 0.5$  
a diffusion constant 
$D_{N} \propto N^{-3/2}$ and a longest autocorrelation time of
$\tau \propto N^{2.5}$ for chains up to 216 monomers on a cubic
lattice. For linear chains, this result is close to ours and also 
close to that of Schweizer's 
``renormalized Rouse model" \cite{schweizer}
which is  intermediate between classical Rouse and reptation
models, predicting $D_{N} \propto N^{-3/2}$ and $\tau \propto N^{5/2}$. 

\bibitem{Skolnick}
A.~Kolinski, J.~Skolnick, R.~Yaris, J. Chem. Phys. {\bf 86}, 1567 (1987);
{\em ibid.} {\bf 84}, 1922 (1986); {\em ibid.} {\bf 86}, 7164 (1987);
{\em ibid.} {\bf 86}, 7174 (1987).

\bibitem{schweizer}
K.~S.~Schweizer, J.~Chem.~Phys. {\bf 91}, 5802; 5822 (1989);
K.~S.~Schweizer and G.~Szamel, J.~Chem.~Phys. {\bf 103}, 1934 (1995).

\bibitem{Cnlinear}
The shape of the correlation functions $C_{n}(t)$ is perhaps
not unique to rings. In their simulations Skolnick et al \cite{Skolnick}
examined for linear chains the survival of long-lived two-chain contacts,
which decayed in a multiexponential manner, with a
significant population at times greater than the longest 
relaxation time $\tau$.

\bibitem{richter}
D.~Richter, L.~Willner, A.~Zirkel, B.~Farago, L.~J.~Fetters, and J.~S.~Huang,
Macromolecules {\bf 37}, 7437 (1994).

\bibitem{reiter91}
J.~Reiter, J. Chem.Phys. {\bf 95}, 1290 (1991).

\end{thebibliography}
\end{document}